# Momentum-Resolved Tunneling Modulation Induced Giant Multistate Resistance in Antiferroelectric Multiferroic Junction

Wei Yang[1,2,†], Yibo Xu[1,2,†], Shen Li[1,2], Jiangchao Han[2], Jiayou Chen[2], Juan-Carlos Rojas-Sánchez[3], Stéphane Mangin[3], Xiaoyang Lin[1,2,*], Weisheng Zhao[1,2]

[1] State Key Laboratory of Spintronics, Hangzhou International Innovation Institute, Beihang University, Hangzhou, 311115, China

[2] Fert Beijing Institute, School of Integrated Circuit Science and Engineering, Beihang University, Beijing, 100191, China

[3] CNRS, Institute Jean Lamour, Université de Lorraine, Nancy F-54000, France

[†]These authors contributed equally

[*]Correspondence and requests for materials should be addressed to:

Xiaoyang Lin: XYLin@buaa.edu.cn

**Abstract:** Multiferroic tunnel junctions (MFTJs), integrating ferroelectric and ferromagnetic functionalities within a single nanoscale device, hold significant promise for non-volatile, multi-state memory and innovative computing paradigms. In conventional MFTJs, tunneling resistance modulation relies primarily on ferroelectric (FE) polarization switching, which alters interfacial electric fields and shifts the Fermi level of adjacent ferromagnetic electrodes. However, achieving high tunnel electroresistance (TER) through this approach demands strong built-in electric fields, which simultaneously hinder FE polarization switching, creating an intrinsic trade-off between reliable data reading and efficient writing. Here, we propose a dual mechanism that combines antiferroelectric (AFE) phase-transition modulation of the evanescent decay states with interfacial spin filtering based on $Fe_3GaTe_2$/bilayer-$In_2Se_3$/$Fe_3GaTe_2$ heterostructure. Beyond altering the electrostatic potential as in AFE–FE switching, the transitions between head-type and tail-type AFE states preserve the centrosymmetric potential profile yet fundamentally modulate the momentum-resolved distribution of evanescent decay rates across the Brillouin zone. When integrated with perfect spin filtering at the $Fe_3GaTe_2$/α-$In_2Se_3$ interface, this mechanism yields a giant TER (~$7.6\times10^3$%), over 4 times that of conventional FE-based MFTJs, and a TMR exceeding $6.8\times10^5$%, enhanced by two orders of magnitude over typical MFTJs. These mechanisms resolve the performance trade-off in MFTJs, enabling six distinct non-volatile resistance states at room temperature.

## Introduction

Multiferroic tunnel junctions (MFTJs) integrate ferroelectric and ferromagnetic functionalities into a single nanoscale device, offering a promising pathway toward non-volatile,[1-3] multi-state memory[4-6] and logic applications.[7-9] In conventional MFTJs, the tunnel electroresistance (TER) effect is primarily driven by ferroelectric (FE) polarization reversal,[10-13] which reverses the interfacial electric fields and shifts the Fermi level in the adjacent ferromagnetic electrodes. As a result, the effective barrier height for electron tunnelling is modulated,[14-16] giving rise to the TER effect, defined as $TER = \frac{G_{max} - G_{min}}{G_{min}}$, where $G_{max}$ and $G_{min}$ denote the maximum and minimum tunnelling conductance, respectively.[17] Simultaneously, the relative magnetic configuration of the electrodes governs the tunnel magnetoresistance (TMR), enabling the coexistence of electric and magnetic degrees of freedom within the same junction. The TMR is quantitatively described by $TMR = \frac{R_{AP} - R_P}{R_P}$, where $R_{P(AP)}$ indicates the resistance for (anti-) parallel configuration of magnetizations.[18, 19]

Generally, achieving a high tunnel electroresistance (TER) requires the presence of a strong interfacial electric field to induce significant modulation of the tunnelling barrier, as shown in Figure 1(a). However, a substantial built-in electric field simultaneously impedes the reversal of ferroelectric (FE) polarization.[20, 21] This intrinsic contradiction poses a fundamental challenge that severely constrains the practical development of multiferroic tunnel junctions. A potential strategy to mitigate this dilemma involves employing two-dimensional symmetric antiferroelectric tunnel junctions (AFTJs), as shown in Figure 1(b&c).[22] In these devices, the anti-ferroelectric switching between antiferroelectric (AFE) and FE states can effectively eliminate internal built-in electric fields[23] and significantly reduce the ferroelectric switching barrier,[24, 25] thus promoting the development of high-performance multiferroic tunnel junctions.[26] However, the TER effect obtained in the proposed symmetric van der Waals MFTJs falls short of the demands for high-density,[27, 28] low-power spintronic devices.[29-31] The reported TER ratio of symmetric vdW MFTJs based on Ag/bilayer-$VSi_2N_4$/Au, $CrSe_2/CuInP_2S_6/CrSe_2$ and $Fe_3GeTe_2$/bilayer-$In_2Se_3/Fe_3GeTe_2$ are only 37.7%, 90% and 744%, respectively[32-34]; which nearly two orders of magnitude lower than TER of conventional asymmetric FTJs.[11, 17, 35, 36]

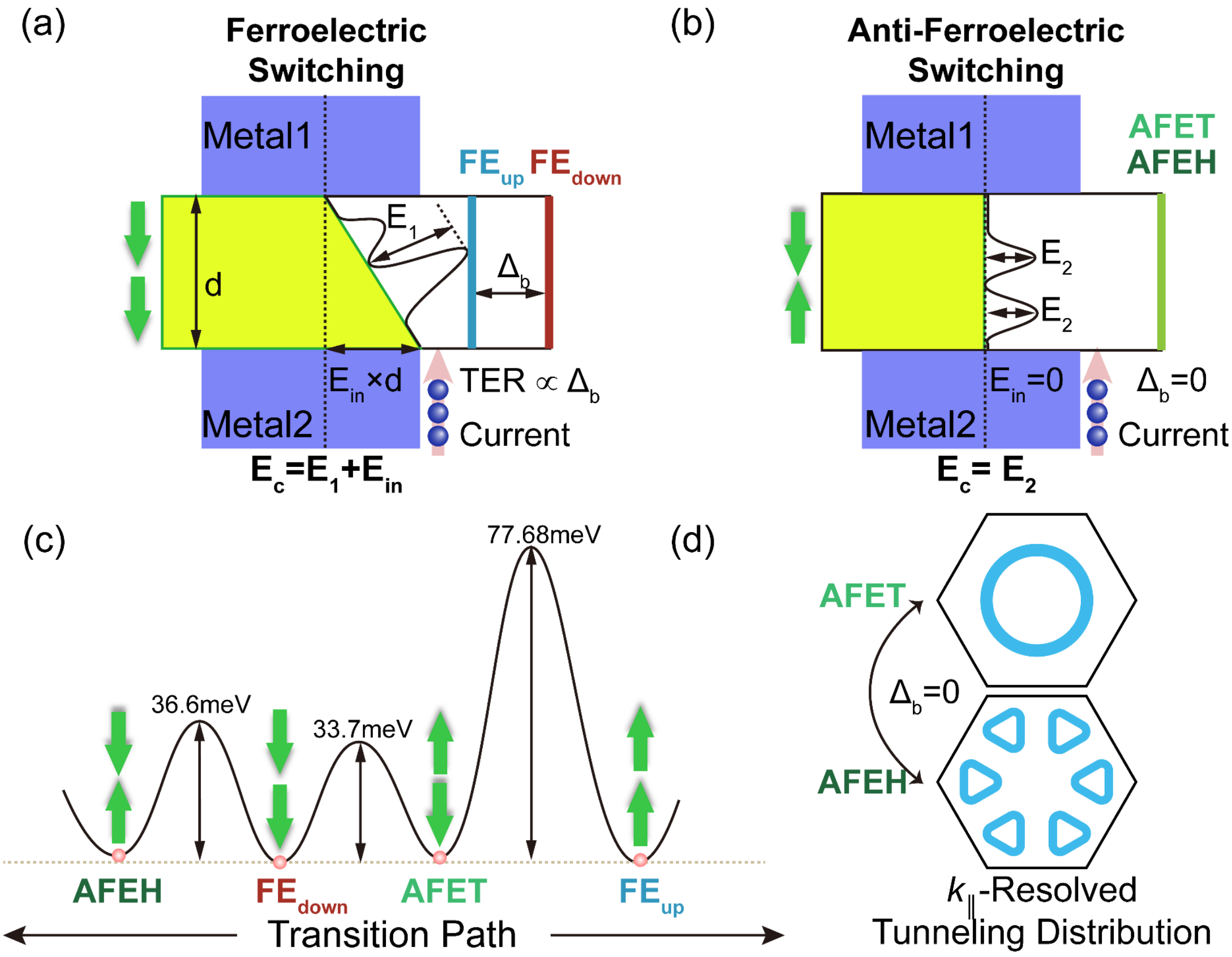

**Figure 1. Schematic Illustration of the Transport Mechanism in High-performance Antiferroelectric Multiferroic Tunnel Junctions (AFE-MFTJ).** (a–b) Comparison between conventional ferroelectric (FE) switching and antiferroelectric (AFE) switching. In FE switching (a), the threshold electric field is $E_c = E_1 + E_{in}$, where $E_1$ is the intrinsic energy barrier for polarization reversal and $E_{in}$ is the built-in electric field arising from the FE polarization. The FE switching modifies the tunneling barrier height by $\Delta_b$, giving rise to the tunnel electroresistance (TER) effect. In contrast, for AFE switching (b), the opposing polarizations in bilayer systems cancel the internal field ($E_{in}$=0), and the switching requires overcoming only the intrinsic barrier $E_2$. As a result, the tunneling barrier height remains nearly unchanged ($\Delta_b$ = 0). (c) Energy landscape of the FE and AFE switching processes, illustrating that the switching barrier 77.68 meV for FE is significantly larger than 36.6meV and 33.7meV for AFE. (d) Momentum-space schematic showing the modulation of tunneling distribution in the Brillouin zone induced by AFE switching between tail-to-tail antiferroelectric (AFET) and head-to-head antiferroelectric (AFEH) states, where the blue region indicates efficient tunneling of electrons. Even in the absence of a barrier height change ($\Delta_b$ = 0), substantial TER can be achieved due to the redistribution of evanescent decay states in momentum space.

In this work, we propose a van der Waals multiferroic tunnel junction (MFTJ) that overcomes the intrinsic trade-off between efficient polarization switching and large

TER, by leveraging a momentum-resolved tunneling modulation mechanism arising from antiferroelectric (AFE) phase transitions, as shown in Figure 1(d). Built upon a $Fe_3GaTe_2$/α-$In_2Se_3$ heterostructure, the device integrates AFE-state–induced redistribution of evanescent decay states in the Brillouin zone with strong spin-filtering effects at the ferromagnet/ferroelectric interface, enabling simultaneous realization of high TER and tunnel magnetoresistance (TMR) without relying on built-in electric fields. The bilayer α-$In_2Se_3$, which is compatible with potential wafer-scale fabrication techniques,[37] exhibits supports both ferroelectric (FE) and antiferroelectric states, and allows multiple AFE configurations characterized by head-to-head (AFEH) and tail-to-tail (AFET) dipole arrangements.[22, 38] Compared to previously studied ferromagnetic electrodes such as $Fe_3GeTe_2$[39] and $Fe_4GeTe_2$,[40, 41] $Fe_3GaTe_2$ exhibits a higher Curie temperature (~380 K) and strong perpendicular magnetic anisotropy $K_u \approx 4.79 \times 10^5$ $J/m^3$ in bulk and $\approx 3.88 \times 10^5$ $J/m^3$ in the 2D limit,[42-44] ensuring robust magnetism at room temperature.

Through first-principles calculations combining density functional theory (DFT) with the non-equilibrium Green's function (NEGF) formalism, we demonstrate that, beyond merely modulating electrostatic potentials, the AFE–FE and AFEH–AFET transitions in bilayer α-$In_2Se_3$ fundamentally modulate the momentum-resolved distribution of evanescent decay rates across the Brillouin zone. This momentum-resovled modulation enables selective control over electron transport pathways, resulting in a high TER exceeding $7.6 \times 10^3$%. Simultaneously, the $Fe_3GaTe_2$/α-$In_2Se_3$ interface exhibits strong spin-filtering effects, leading to a substantial tunnel magnetoresistance (TMR) exceeding $10^5$%. Our findings establish AFE phase transitions, not only between FE and AFE states but also within distinct AFE configurations, as a useful mechanism to simultaneously achieve high TER and TMR, marking a significant step toward next-generation multifunctional electronics.

## Results and Discussion

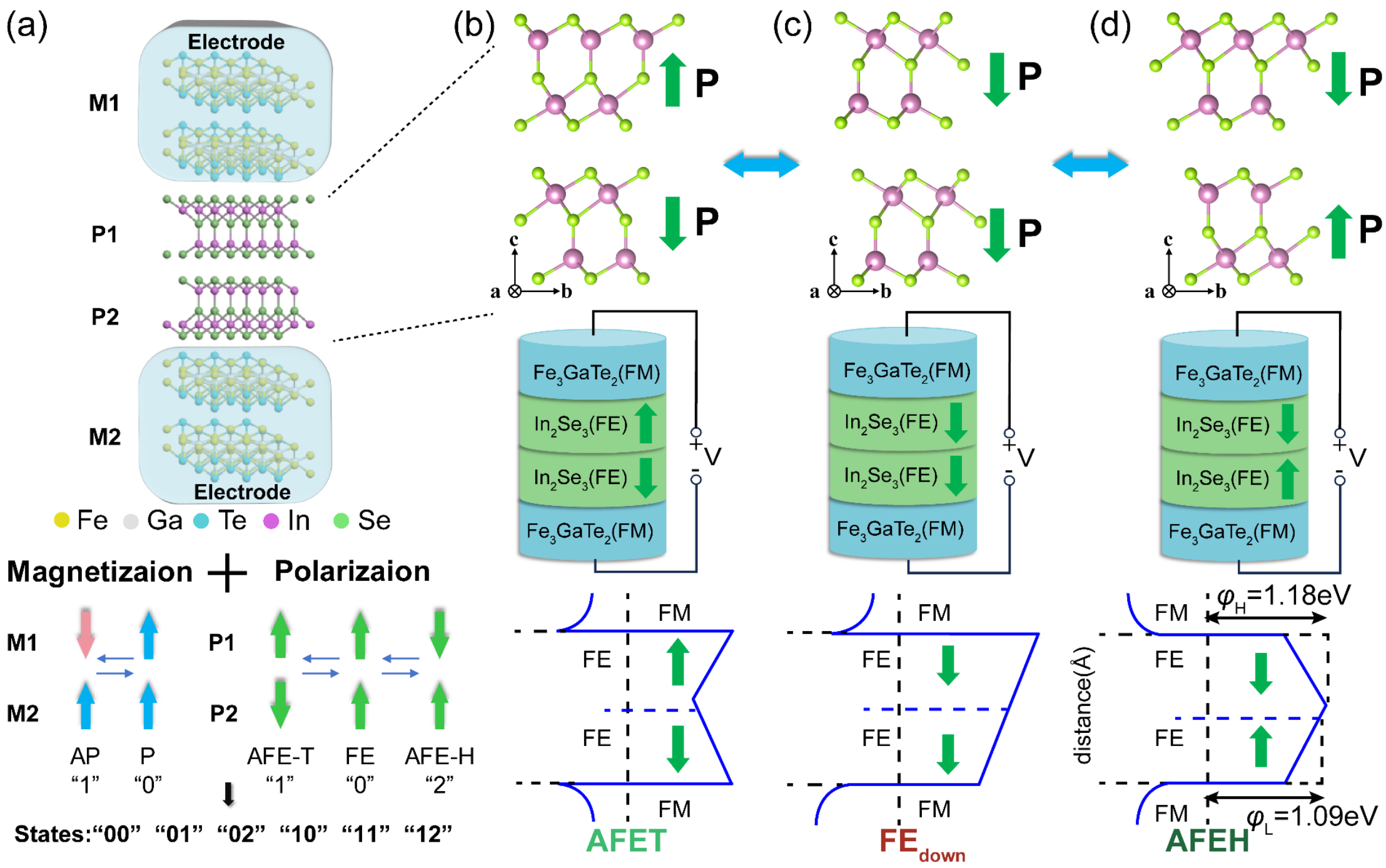

**Figure 2. Multi-State Storage Concept Based on AFE-MFTJ.** (a) Atomic structure of the proposed device, where the blue regions denote two semi-infinite $Fe_3GaTe_2$ electrodes. M1(2) and P1(2) represent the magnetization directions of the ferromagnetic layers and the polarization orientations of the ferroelectric layers, respectively. The combined configurations of magnetic and electric order parameters yield six distinct non-volatile resistance states. (b–d) Top panels: side views of the bilayer α-$In_2Se_3$ in the (b) AFET, (c) ferroelectric (FE), and (d) AFEH polarization states. Green arrows indicate the polarization directions of individual α-$In_2Se_3$ monolayers. Bottom panels: corresponding electrostatic potential profiles across the tunneling barrier for each polarization state, illustrating their distinct symmetry and barrier modulation characteristics.

We investigated the spin transport properties of the proposed device using a classical two-terminal model. As illustrated in Figure 2(a), the junction consists of two semi-infinite $Fe_3GaTe_2$ electrodes and a central scattering region composed of a $Fe_3GaTe_2$/bilayer-α-$In_2Se_3$/$Fe_3GaTe_2$ heterostructure. The α-$In_2Se_3$ and $Fe_3GaTe_2$ exhibit lattice constants of 4.063 Å and 3.986 Å respectively, resulting in a minimal lattice mismatch of 1.97%. Consequently, a 1×1 unit cell was adopted for both materials in constructing the device model. When $Fe_3GaTe_2$ serves as the substrate, the slight tensile strain imposed on the overlying α-$In_2Se_3$ layers exerts a negligible influence on their electronic structure. The excellent compatibility indicates a high degree of experimental feasibility for fabricating such tunnel junctions.

The bilayer α-$In_2Se_3$ barrier adopts different polarization configurations depending on the orientation of the ferroelectric polarization in each layer, yielding AFET, AFEH, and FE states. The atomic stacking structures corresponding to these configurations are illustrated in top panels of Figure 2(b-d), respectively. To ensure the feasibility of the proposed six magnetoelectric states, we evaluated both their structural stability and non-volatility. Binding energy calculations confirm that all configurations are thermodynamically stable (Tables S3 and S7). The ferroelectric switching barriers and perpendicular magnetic anisotropy energies (Tables S5 and S4) are sufficiently high to prevent spontaneous reversal, indicating reliable non-volatile behavior. Moreover, the intrinsic ferroelectricity of α-$In_2Se_3$ and robust ferromagnetism of $Fe_3GaTe_2$ have been experimentally demonstrated at room temperature,[42, 45, 46] supporting the stability and functionality of the heterostructure in practical applications.

Unlike conventional magnetic tunnel junctions, where the asymmetry in the work function between the top and bottom surfaces of the ferroelectric barrier induces substantial built-in electric fields that hinder polarization switching (Table S6), the incorporation of antiferroelectric materials can mitigate this problem. As illustrated in Figure 2 (b&d), the electrostatic potential profiles for the AFET and AFEH states demonstrate that the surface potentials at the ferromagnetic electrodes remain symmetric, thereby eliminating internal electric fields while retaining the capacity to modulate the barrier height, akin to ferroelectric switching.

The proposed device architecture facilitates reversible switching among six non-volatile memory states through controlled application of external voltage, as shown in Figure 2(b-d) and Table S5. As experimentally demonstrated by the switching of ferroelectric polarization through an out-of-plane electric field,[47-49] the application of external voltages exceeding the critical switching thresholds enables reversible transitions between antiferroelectric and ferroelectric states, thereby directly influencing spin-dependent transport. To evaluate the feasibility of polarization switching, we calculated the critical electric fields by $E = \frac{\Delta E_{barrier}}{|P_{z,1} - P_{z,2}|}$ (Tables S5),[50] and the calculated values are consistent with those observed experimentally.[51] On the other hand, the efficient magnetization switching of $Fe_3GaTe_2$ can be achieved by the application of an external magnetic field application[52] or via spin–orbit torque (SOT).[53-55] Recent theoretical studies indicate that electric-field–driven reversal of ferroelectric polarization can deterministically regulate the magnetic ground state of adjacent ferromagnetic layers.[30, 56, 57] In addition, the impact of polarization switching on the topological states of adjacent ferromagnetic layers has been elucidated, providing critical insight into the fundamental mechanisms of magnetoelectric coupling.[58, 59] Experimentally, achieved in electrically controlling of ferromagnetic moments has been via magnetoelectric coupling,[60, 61] which offers a powerful pathway for magnetization switching in the proposed device architecture, enabling comprehensive electrical control of both ferroelectric and magnetic orders without requiring additional magnetic fields.

Once the ferroic states are written, the device allows non-destructive electrical readout through TER and TMR. To investigate the transport behavior, we calculated the spin-resolved transmission spectra, as shown in Supplementary Figure S1(a). The results reveal that, irrespective of whether the barrier is in the ferroelectric or antiferroelectric state, the spin-up and spin-down transmission probabilities at the Fermi level differ by nearly four orders of magnitude when the ferromagnetic electrodes adopt a parallel magnetic configuration. In contrast, under the antiparallel configuration (Figure S1(b)), the difference in transmission between spin-up and spin-down channels is significantly reduced.

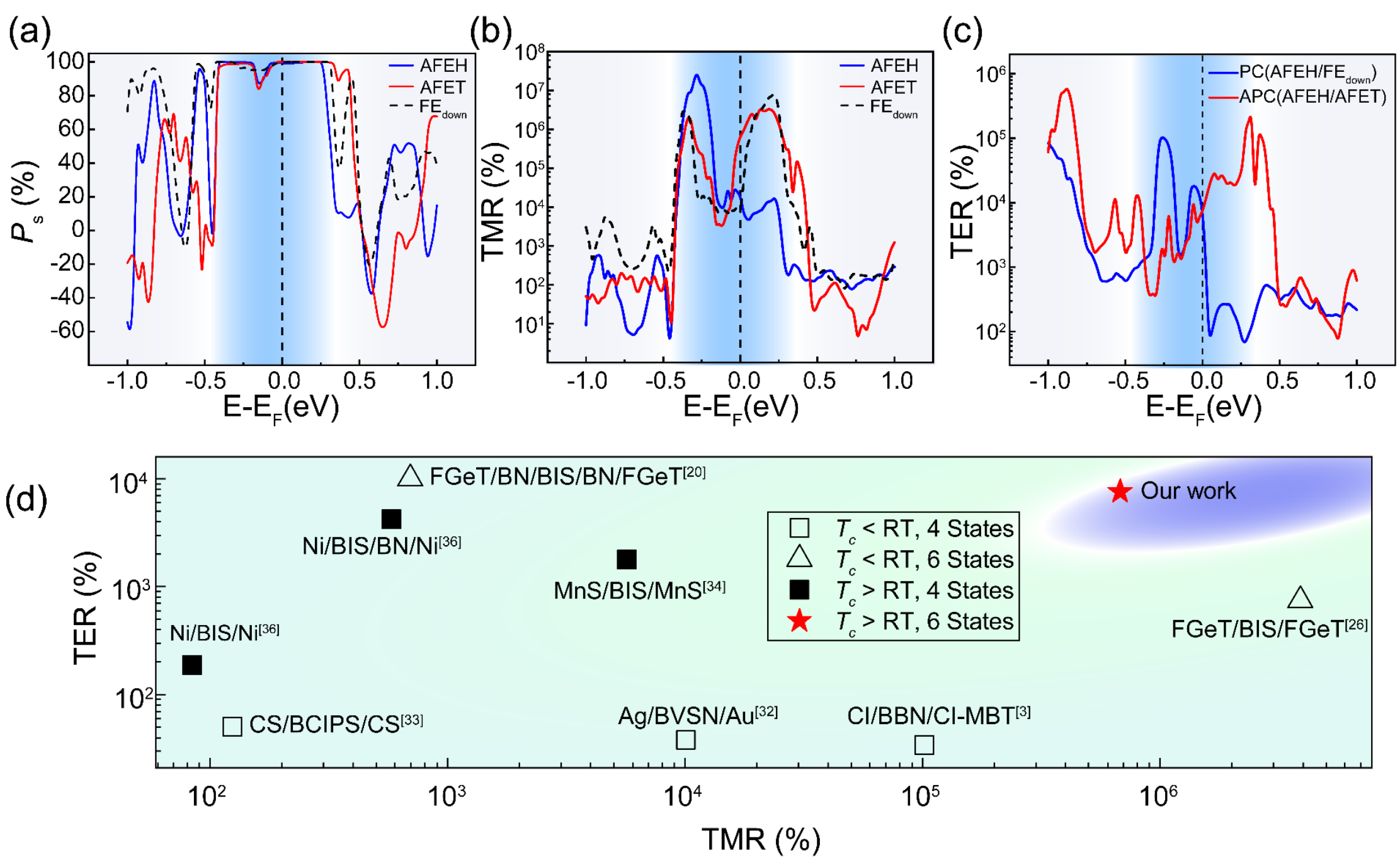

**Figure 3. Spin Transport in $Fe_3GaTe_2$/bilayer-$In_2Se_3$/$Fe_3GaTe_2$ AFE-MFTJ.** (a) Spin polarization spectra of different polarization states under the parallel configuration of ferromagnetic electrodes. The blue regions indicate areas of high spin polarization (~100%). (b) TMR spectra in devices with different ferroelectric polarization states. The dashed line represents the ferroelectric polarization state, while the solid line represents the antiferroelectric polarization state. (c) TER spectra in devices with different ferromagnetic electrode magnetization configurations, where the blue area represents the band gap region of the bilayer ferroelectric. For the PC magnetic configuration, the TER ratio is obtained from the calculations between AFEH and $FE_{down}$, whereas for the APC magnetic configuration, the TER ratio is obtained from the calculations between AFEH and AFET. (d) Comparative Performance Landscape of MFTJ (Table S2) with Ref.3, 20, 26, 32, 33, 34, 36. Hollow and solid represent MFTJ devices with Curie temperatures ($T_C$) below and above room temperature (RT), respectively. The two axes correspond to TER and TMR, and the shape of the icon correspond the number of distinguishable resistance states, enabling a comprehensive comparison of multifunctional performance.

Subsequently, we evaluated the spin polarization of the device under the parallel magnetic configuration, calculated as $P_s = \frac{T_{PC}^{\uparrow} - T_{PC}^{\downarrow}}{T_{PC}^{\uparrow} + T_{PC}^{\downarrow}}$,[62] where $T_{PC}^{\uparrow}$ and $T_{PC}^{\downarrow}$ denote the spin-up and spin-down transmission coefficients, respectively. As shown in Figure 3(a), the computed spin polarization approaches nearly 100% within the bandgap range of bilayer $In_2Se_3$ (marked by the blue shaded region), indicating that all polarization states of the ferroelectric barrier exert significant influence on spin filtering.

Table 1 summarizes the calculated conductance values at the Fermi level for all resistance states in the AFE-MFTJ. Owing to the exceptionally high spin polarization achieved in this architecture, the device exhibits impressive TMR ratios of $2.05 \times 10^4$%, $6.79 \times 10^5$%, and $2.29 \times 10^5$% under the AFEH, AFET, and FE polarization states, respectively. These remarkable values are three orders of magnitude larger than that in the $Mn_2Se_3/In_2Se_3$ heterostructures.[32]

Moreover, these high TMR values (> $10^4$%) remain robust across a broad energy range, as indicated by the blue region in Figure 3(b), significantly mitigating the sensitivity of device performance to variations in applied bias voltage and environmental fluctuations, thereby ensuring stable and reliable operation under diverse conditions.

Table 1. Calculated conductance (in $e^2/h$), TMR, and TER ratio of the $Fe_3GaTe_2$/bilayer-$In_2Se_3$/$Fe_3GaTe_2$ van der Waals antiferroelectric multiferroic tunnel junction at the Fermi level.

| | PC | APC | **TMR (%)** |
|---|---|---|---|
| AFEH | $9.67\times10^{-3}$ | $4.68\times10^{-5}$ | $2.05\times10^{4}$ |
| AFET | $4.29\times10^{-3}$ | $6.32\times10^{-7}$ | $6.79\times10^{5}$ |
| $FE_{down}$ | $1.26\times10^{-4}$ | $1.82\times10^{-6}$ | $6.86\times10^{3}$ |
| **TER (%)** | 7574.6% | 7305.06% | |

Table 1 further reveals that the conductance of the AFEH state consistently surpasses that of both the AFET and FE states in both parallel and antiparallel magnetic configurations. This leads to a pronounced TER of approximately $7\times10^3$% in the $Fe_3GaTe_2$/bilayer-$In_2Se_3$/$Fe_3GaTe_2$ junction, a value that notably exceeds the TER typically attainable via conventional ferroelectric polarization reversal mechanisms in conventional MFTJs, for example, it is ~4 times higher than the ferroelectric switching induced TER of $Mn_2Se_3/In_2Se_3$ heterostructures.[32] Both theoretical[22] and experimental[49] investigations have demonstrated the emergence of tilted polarization metastable states during switching processes. These metastable states manifest as switching curves similar to those shown in Figure S13, yet they do not compromise the stability of the initial and final polarization states. Subsequent transport calculations confirm that ferroelectric domain walls formed during switching have a negligible effect on both TER and TMR. More importantly, the simultaneous realization of both high TMR and high TER in this device substantially outperforms previously reported systems, as benchmarked in Figure 3(d), firmly positioning this architecture as a promising candidate for next-generation high-density, low-power spintronic applications.

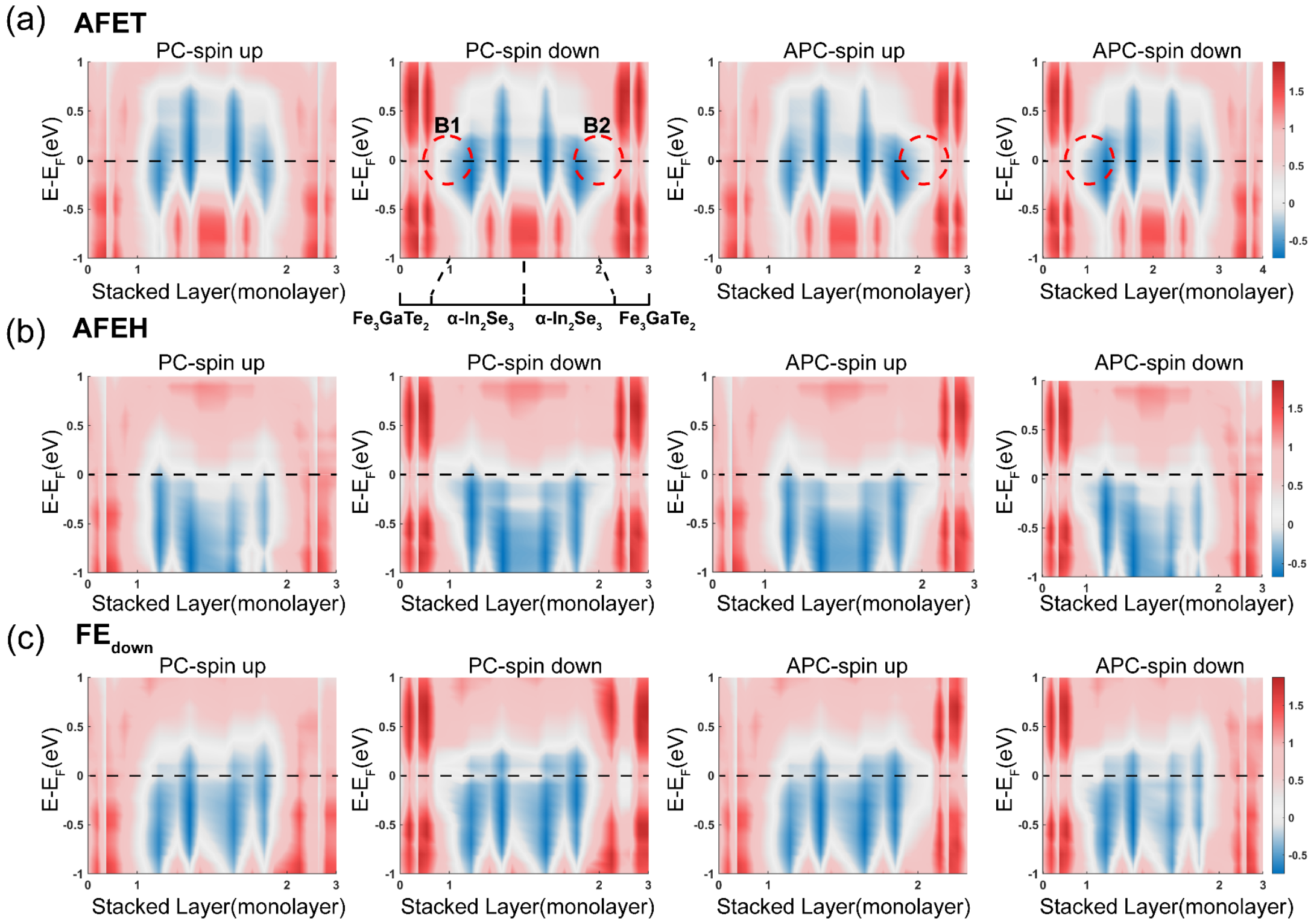

**Figure 4. Local Density of States (LDOS) Along the Transport Direction.** (a)-(c) the LDOS, of tail-to-tail antiferroelectric (AFET), head-to-head antiferroelectric (AFEH), and ferroelectric down ($FE_{down}$) respectively. The *x*-axis represents the number of stacked layers of the tunnel junction in units of monolayer atoms, from left to right: $Fe_3GaTe_2$, α-$In_2Se_3$, $Fe_3GaTe_2$. The four sub-figures from left to right are: parallel configuration spin up and spin down, antiparallel configuration spin up and spin down. The black dashed line represents the Fermi level.

To further elucidate the physical mechanisms underlying the remarkably high TMR, TER, and spin polarization observed in this device, we performed calculations of the real-space local density of states (LDOS) along the transport direction to get the insight into the spin-dependent tunnelling processes occurring within the junction. As illustrated in Figure 4, the four subpanels present the LDOS distributions for the spin-up and spin-down channels under parallel and antiparallel magnetic configurations, respectively. In the FE polarization state, the potential barrier height decreases progressively along the direction of polarization. This gradient reflects the presence of a built-in electric field within the ferroelectric tunnelling layer. By contrast, in AFE states, this internal field is effectively compensated due to the opposing polarizations in the two $In_2Se_3$ layers, resulting in a centrosymmetric barrier profile.

Furthermore, under all polarization states, the spin-up and spin-down LDOS distributions exhibit nearly identical profiles in the antiparallel magnetic configuration, indicating suppressed spin polarization in this configuration. Consequently, our subsequent analysis focuses on the spin-up LDOS in both parallel and antiparallel configurations to capture the key spin-transport features. Taking the AFET polarization state as an example, as shown in Figure 4(a), two pronounced barriers (B1 and B2) emerge in the spin-down transport channel under the parallel magnetic configuration, while no significant barriers are present in the spin-up channel. This spin-dependent asymmetry accounts for the substantial difference in transmission coefficients between spin-up and spin-down electrons under parallel alignment. By contrast, in the antiparallel configuration, both spin-up and spin-down channels exhibit a single barrier, leading to ignorable difference in transmission between the spin channels in this state.

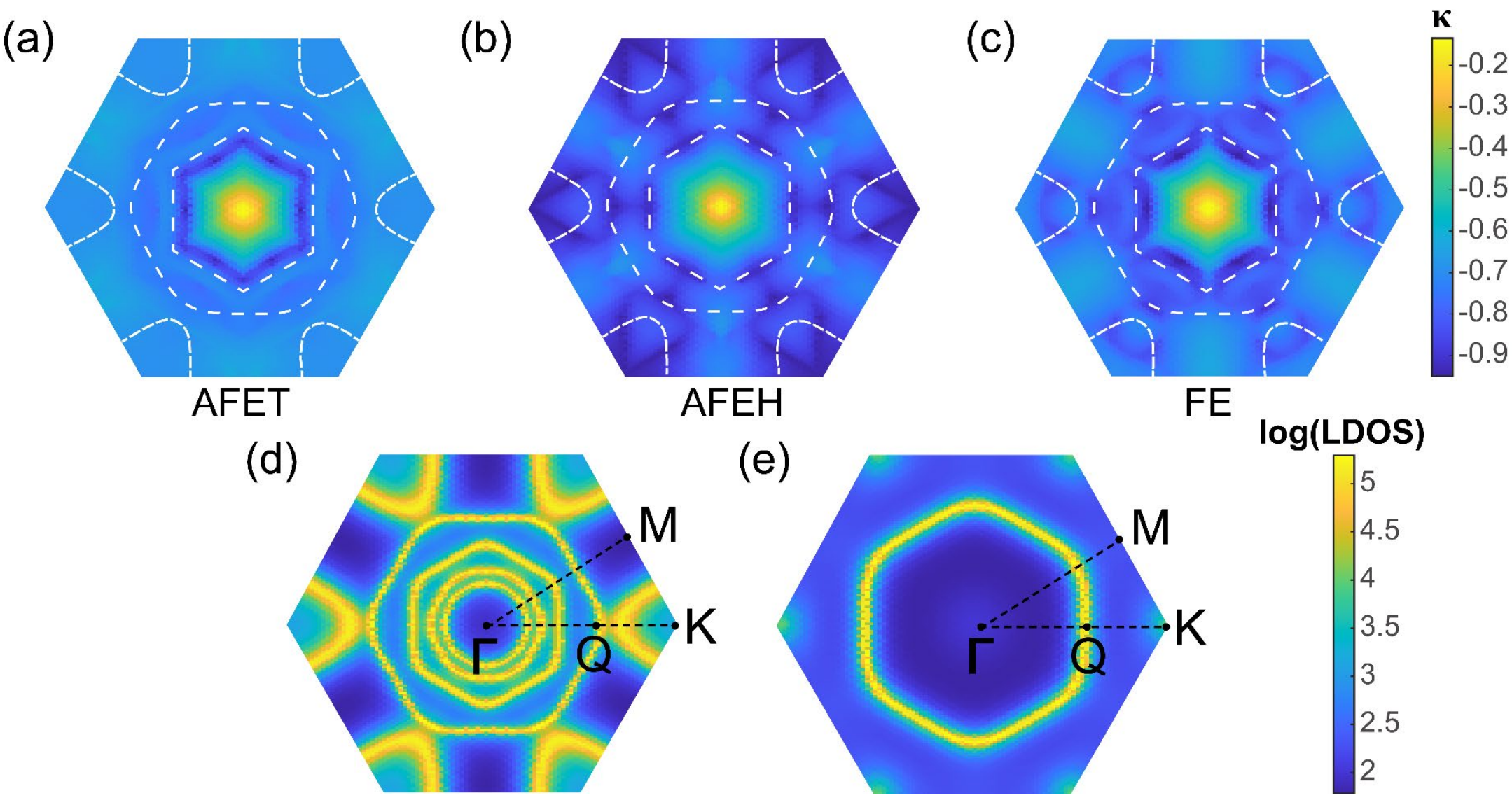

**Figure 5. AFE-Driven Modulation of Tunneling Channels and Spin-Resolved Filtering.** (a)-(c) are the k-dependent minimum decay rate distributions of the tail-to-tail antiferroelectric (AFET), head-to-head antiferroelectric (AFEH) and ferroelectric (FE) polarization states. (d) and (e) are the density of states distributions of spin-up and spin-down $Fe_3GaTe_2$. The white dashed line enclosed area of (a)-(c) indicates the high density of states distribution region of spin-up $Fe_3GaTe_2$.

Moreover, we observe that in the parallel magnetic configuration, switching from FE-up to FE-down polarization yields a TER as high as 2320%, which can be understand as the significant change in barrier height driven by interfacial electric field reversal[11, 14, 63, 64](see Table S9). Remarkably, in antiparallel magnetic configuration, even the transitions between the AFEH and AFET states produce a pronounced TER reaching 7305%, demonstrating that antiferroelectric switching can achieve TER values significantly exceeding those typical of conventional ferroelectric polarization reversal. This finding highlights a crucial advantage of antiferroelectric tunnel junctions: unlike ferroelectric switching, which relies primarily on interfacial electric field reversal, AFE phase transitions modulate the tunnelling barrier through changes in the momentum-resolved distribution of evanescent decay rates, offering an additional and highly effective mechanism for tuning TER. The underlying mechanism can be explained as follows: in the parallel magnetic configuration, the strong spin-selective barrier at the ferroelectric/ferromagnetic interface causes spin-up electrons to overwhelmingly dominate the transport process, thereby reducing the impact of barrier modulation induced by switching between AFET and AFEH polarization states. In contrast, under the antiparallel magnetic configuration, both spin channels are effectively suppressed by their respective spin-selective barriers. In this regime, the total transmission becomes highly sensitive to variations in the barrier attenuation rates caused by transitions between AFEH and AFET states, enabling the realization of exceptionally high TER.

To elucidate this phenomenon, we computed the dependence of the minimum decay constant κ on the in-plane wavevector $k_{\|}$ for different polarization states of bilayer α-$In_2Se_3$, as shown in Figure 5 (a–c). The decay constant is defined through the complex wavevector $k_z = k + i\kappa$, where the imaginary part κ represents the exponential attenuation of evanescent states. The corresponding wavefunction can be expressed as $\sim e^{-k_z}$.[65, 66] A smaller absolute value of κ indicates more efficient tunnelling of the associated electronic states.

We further calculated the spin-resolved $k_{\|}$-dependent density of states (DOS) in $Fe_3GaTe_2$, which determines the density of incident states. The spin-up DOS is predominantly distributed around the K point, forming a triangular region centered at K and a hexagonal region between the Γ and Q points (Figure 5(d)). In contrast, the spin-down DOS is largely confined to a narrow circular region near the K point and a hexagonal path traversing the Q point (Figure 5(e)). Notably, the DOS at the Fermi level reveals an intrinsic spin polarization in $Fe_3GaTe_2$, favoring the spin-up channel. Crucially, the regions of high spin-up DOS coincide with areas in α-$In_2Se_3$ exhibiting low decay constants across different polarization states, facilitating efficient tunnelling of spin-up electrons through the barrier. Conversely, the regions of high spin-down DOS correspond to higher decay constants, severely suppressing spin-down electron transmission and giving rise to the spin-dependent barriers B1 and B2. This selective tunnelling mechanism enhances spin filtering and enables high TMR and nearly perfect spin polarization. Moreover, significant variations in the decay constant are observed near the triangular region around the Q point, corresponding to the spin-degenerate region of $Fe_3GaTe_2$, across different polarization states. These differences lead to pronounced changes in the total spin-resolved transmission coefficients, thereby producing substantial TER in the device. This interpretation is further supported by our complex band structure calculations of the FGT/α-$In_2Se_3$/FGT junction under both magnetic configurations (Figures S9 and S10), which reveal pronounced shifts in the imaginary part of the wavevector across the AFE phase transition and confirm that the AFE-induced modulation of evanescent states is the key mechanism governing transmission in our device.

Collectively, these two mechanisms—the spin-selective matching between $Fe_3GaTe_2$ electronic states and α-$In_2Se_3$ decay profiles, as well as the polarization-induced modulation of decay constants—allow the device to simultaneously achieve high TMR and large TER. This significantly elevates the overall performance of multiferroic tunnel junctions and underscores the unique advantages of the fully two-dimensional van der Waals heterostructure architecture in designing high-performance spintronic devices.

## Conclusion

In summary, we have proposed a multiferroic tunnel junction architecture that integrates bilayer antiferroelectric α-$In_2Se_3$ with high–Curie temperature ferromagnet $Fe_3GaTe_2$, achieving exceptional performance at room temperature. Unlike conventional MFTJs that rely on built-in electric fields and polarization-induced band bending, our device exploits purely AFE-to-AFE phase transitions to modulate the tunneling process through changes in the momentum-resolved evanescent decay spectrum—without altering the electrostatic barrier height. This mechanism, combined with perfect spin filtering at the $Fe_3GaTe_2$/α-$In_2Se_3$ interface, enables simultaneous realization of an ultra-high tunnel magnetoresistance exceeding $6.8\times10^5$% and a giant tunnel electroresistance of ~$7\times10^3$%. This approach resolves the trade-off in multiferroic tunnel junctions between achieving high TER for reliable readout and maintaining efficient polarization switching for low-power write operations. Thanks to the robust ferroic orders, the combination of three electric polarization states and two magnetic configurations leads to six non-volatile resistance states accessible at room-temperature.

## Method

The device structure is optimized using density functional theory (DFT) in the Vienna ab initio simulation package (VASP).[67, 68] The generalized gradient approximation (GGA) Perdew-Burke-Ernzerhof (PBE) functional is employed to describe the exchange-correlation interactions during the calculations.[69, 70] We systematically studied different Hubbard U values (Table S10, Figures S11, S12), and found that U = 0 yields the best agreement with both experimental results[71] and prior theoretical studies.[72-74] A cut-off energy of 450 eV is adopted for the plane wave basis in the structural optimization, and the Brillouin zone was sampled with a Γ-centered Monkhorst-Pack grid of 12 × 12 × 1 k-points. The convergence criterion for forces is set to 0.02 eV·$Å^{-1}$. To avoid periodicity effects during the simulation, a vacuum layer of 20 Å is built in the z-direction. The DFT-D3 method was employed to describe the van der Waals interactions between layers.[75, 76] The electronic transport properties were calculated using the Nanodcal package, which combines density functional theory (DFT) with the non-equilibrium Green's function (NEGF) method.[77, 78] In this approach, physical quantities are expanded using a linear combination of atomic orbital (LCAO) basis sets at the double-zeta plus polarized (DZP) level.[79] For the electronic transport calculations, the cutoff energy is set to 160 Rydberg, and the convergence criteria for both the density matrix and the Hamiltonian matrix are established at $1 \times 10^{-4}$ eV. The k-point grids for spin-dependent transport calculations are configured as 150 × 150 × 1, 16 × 16 × 1 and 150 × 150 × 1, respectively. These grids were used to compute the local density of states (LDOS) of the FM electrode $Fe_3GaTe_2$ and the evanescent states of the barrier layer α-$In_2Se_3$. Under equilibrium conditions, the transmission coefficients were averaged over the Brillouin zone:[78]

$$T_\sigma(E) = \frac{1}{A_{\mathrm{BZ}}} \int_{\mathrm{BZ}} T_\sigma(E, k_\parallel) \mathrm{d}k_\parallel \tag{1}$$

where $A_{\mathrm{BZ}}$ is the area of the Brillouin zone, and $T_\sigma(E, k_\parallel)$ is the transmission coefficient for spin σ at the transverse Bloch wavevector $k_\parallel$ and energy $E$. TMR is a key parameter for evaluating magnetic devices, as it reflects the strength of a device's information readout capability. Increasing the TMR can improve the signal-to-noise ratio of the device. TMR is calculated using the following formula:[80]

$$TMR = \frac{[(T_{PCup} + T_{PCdown}) - (T_{APCup} + T_{APCdown})]}{T_{APCup} + T_{APCdown}} \times 100\% \tag{2}$$

where $T_{PCup}$ and $T_{PCdown}$ are the transmission coefficients for spin-up and spin-down in the parallel configuration, respectively. And $T_{APCup}$ and $T_{APCdown}$ are the transmission coefficients for spin-up and spin-down in the anti-parallel configuration, respectively. TER under the same equilibrium state is calculated using the following formula:[81]

$$TER = \frac{G_{max} - G_{min}}{G_{min}} \times 100\% = \frac{T_{max} - T_{min}}{T_{min}} \times 100\% \tag{3}$$

where $T_{max}$ and $T_{min}$ represent the maximum and minimum values of the transmission coefficients corresponding to the ferroelectric polarization states AFEH, AFET, and FE, under parallel and antiparallel magnetic configurations of the ferromagnetic electrodes, respectively.

## Conflicts of interest

There are no conflicts to declare.

## Acknowledgements

This work was supported by the National Natural Science Foundation of China (T2394475, 62371019, and 52261145694), the Beijing Natural Science Foundation (No. 4232070), the Research Start-up Funds of Hangzhou International Innovation Institute of Beihang University (2024KQ052 and 2025BKZ001), and the International Mobility Project (No. B16001). and the postgraduate research opportunities program of HZWTECH (HZWTECH-PROP). We gratefully acknowledge HZWTECH for providing computation facilities.

## Author contributions

Yang W and Xu Y proposed the idea and performed the calculations. Li S, Han J, Chen J, Rojas J, Mangin S and Lin X participated in the analysis of the data. Lin X supervised the project. All authors contributed to the general discussion.

## Supporting Information

The Supporting Information is available free of charge at xxxxx.

Detailed electric band structure and transmission spectrum of each state; transport properties of the upward-polarized ferroelectric device; the atomic model and performance evaluation of ferroelectric domain wall; complex band structures of each state; magnetic moments, density of states, and band structures of $Fe_3GaTe_2$ under different U values; free energy and binding energy of each of the six states; surface potential and magnetoelectric coupling energy of each state; out-of-plane electric dipole moments of the four ferroelectric polarization states; magnetic moment energy barriers; transition paths and energy barriers between the FE and AFE states.